\documentclass{comjnl}

\usepackage{amsmath}
\usepackage{cite}
\usepackage{amsmath,amssymb,amsfonts}
\usepackage{algorithmic}
\usepackage{graphicx}
\usepackage{textcomp}
\usepackage{xcolor}
\usepackage{booktabs}
\usepackage{multirow}
\usepackage{array}
\usepackage[colorlinks=true, linkcolor=black, citecolor=black, urlcolor=black]{hyperref}
\usepackage{tikz}
\usetikzlibrary{positioning,arrows,calc}

\newcommand{\model}{DynMM-Explain-LLMRec}

\begin{document}

\title[Bridging Collaborative Filtering and Large Language Models]{Bridging Collaborative Filtering and Large Language Models with Dynamic Alignment, Multimodal Fusion and Evidence-grounded Explanations}
\author{Bo Ma}
\affiliation{Department of Software \& Microelectronics, 
Peking University, Beijing, China} \email{ma.bo@pku.edu.cn}
\author{LuYao Liu}
\affiliation{Civil, Commercial and Economic Law School, 
China University of Political Science and Law, Beijing, China} \email{luyaoliu661@gmail.com}
\author{Simon Lau}
\affiliation{Financial Media, 
Peking University, ChangSha, China} \email{liuximing1995@gmail.com}
\author{Chandler Yuan}
\affiliation{Department of Software \& Microelectronics, 
Peking University, Beijing, China} \email{chandleryuan2019@163.com}
\author{XueY Cui}
\affiliation{Department of Software \& Microelectronics, 
Peking University, Beijing, China} \email{xueyangcuipku@163.com}
\author{Rosie Zhang}
\affiliation{Department of Software \& Microelectronics, 
Peking University, Beijing, China} \email{rosiezhang2012@163.com}

\shortauthors{B. Ma et al.}

\received{00 January 2009}
\revised{00 Month 2009}

\keywords{Recommendation systems; language models; collaborative filtering; adaptive learning; multimodal integration; interpretable AI}

\begin{abstract}
Recent research has explored using Large Language Models for recommendation tasks by transforming user interaction histories and item metadata into text prompts, then having the LLM produce rankings or recommendations. A promising approach involves connecting collaborative filtering knowledge to LLM representations through compact adapter networks, which avoids expensive fine-tuning while preserving the strengths of both components. Yet several challenges persist in practice: collaborative filtering models often use static snapshots that miss rapidly changing user preferences; many real-world items contain rich visual and audio content beyond textual descriptions; and current systems struggle to provide trustworthy explanations backed by concrete evidence. Our work introduces \model{}, a framework that tackles these limitations through three key innovations. We develop an online adaptation mechanism that continuously incorporates new user interactions through lightweight modules, avoiding the need to retrain large models. We create a unified representation that seamlessly combines collaborative signals with visual and audio features, handling cases where some modalities may be unavailable. Finally, we design an explanation system that grounds recommendations in specific collaborative patterns and item attributes, producing natural language rationales users can verify. Our approach maintains the efficiency of frozen base models while adding minimal computational overhead, making it practical for real-world deployment.
\end{abstract}

\maketitle

\section{Introduction}

Language models have emerged as powerful tools for recommendation tasks, processing user interaction histories and item descriptions to generate personalized suggestions \cite{s3rec,lightgcn}. A particularly effective strategy involves bridging collaborative filtering techniques with language model representations through compact projection layers, which preserves computational efficiency while maintaining recommendation quality. However, several practical obstacles limit widespread adoption: First, collaborative filtering models typically rely on fixed training snapshots that struggle to capture evolving user preferences or trending content. Second, modern recommendation scenarios involve rich multimedia content---images, videos, and audio---that purely text-based approaches cannot fully leverage. Third, users increasingly demand transparent recommendations with clear justifications, yet current systems often function as black boxes without interpretable reasoning.

This paper introduces \textbf{\model{}}, a framework that addresses these challenges while maintaining the efficiency of frozen base models. Our approach centers on three complementary techniques: First, we develop an adaptive alignment mechanism that continuously learns from new user interactions through lightweight modules, enabling real-time adaptation without expensive model retraining. Second, we create a unified multimodal representation that seamlessly integrates collaborative patterns with visual and audio content, providing robust performance even when certain modalities are missing. Third, we implement an evidence-based explanation system that grounds recommendations in specific user behaviors and item characteristics, generating natural language justifications that users can understand and verify. Figure~\ref{fig:framework} illustrates our overall architecture. The key contributions of this work include:
\begin{itemize}
  \item A lightweight online adaptation strategy that keeps pace with changing user preferences through small, continuously updated modules while preserving the efficiency of frozen base models.
  \item A flexible multimodal integration approach that combines collaborative signals with visual and audio features through shared representations, improving performance particularly for items with limited interaction history.
  \item An interpretable recommendation mechanism that provides evidence-backed explanations by identifying relevant collaborative patterns and item attributes, with built-in safeguards against spurious reasoning.
  \item Comprehensive evaluation covering temporal adaptation, cold-start scenarios, computational efficiency, and explanation quality, demonstrating practical viability for production systems.
\end{itemize}

\section{Related Work}
\textbf{Collaborative filtering developments.} Modern collaborative filtering has benefited from graph neural networks (LightGCN \cite{lightgcn}) and self-supervised techniques (SGL \cite{sgl}). Sequential models such as S3-Rec \cite{s3rec} capture how user preferences evolve over time. \textbf{Language models in recommendation.} Researchers have investigated using language models for recommendation through prompt engineering (P5 \cite{p5}, ChatRec \cite{chatrec}) and representation alignment techniques that connect collaborative knowledge with language model embeddings (TALLRec \cite{tallrec}, A-LLMRec \cite{allmrec}). CLLM4Rec \cite{cllm4rec} explores deeper integration strategies. \textbf{Multimodal approaches.} Systems like MMGCN \cite{mmgcn} and LATTICE \cite{lattice} leverage visual content alongside traditional signals. Pre-trained encoders such as CLIP \cite{clip} and wav2vec2 \cite{wav2vec2} enable effective multimodal representation learning. \textbf{Adaptive systems.} Streaming recommendation methods \cite{streamingrec} address the challenge of evolving user preferences and item catalogs. \textbf{Interpretable recommendations.} Growing interest in explanation generation has led to surveys \cite{explainrec} and practical methods \cite{narre,explain} that aim to make recommendation decisions more transparent and trustworthy.

\section{Preliminaries and Overview}
Let \(\mathcal{U}\) and \(\mathcal{I}\) denote users and items. A CF backbone maps a user \(u\) and item \(i\) to embeddings \(\mathbf{r}_u, \mathbf{e}^{cf}_i \in \mathbb{R}^{d}\). A text encoder yields \(\mathbf{e}^{txt}_i \in \mathbb{R}^{d_t}\), while an LLM has token dimension \(d_\ell\). Alignment-based recommenders learn small networks to (i) fuse CF and side information into a joint latent \(\mathbf{z}_i\), and (ii) project user/item signals into the LLM token space as soft prompts, keeping CF/LLM frozen. The LLM receives a prompt containing a user representation, history summaries, and candidate item tokens to produce rankings or generations.

We assume a frozen \textit{base aligner} that maps \((\mathbf{e}^{cf}_i, \mathbf{e}^{txt}_i)\) to a joint item latent \(\mathbf{z}^{base}_i\) and a user latent \(\mathbf{h}^{base}_u\). Our goal is to enhance this pipeline with dynamic updates, multimodal fusion, and faithful explanations while preserving efficiency.

\begin{figure}[!t]
\centering
\resizebox{\columnwidth}{!}{%
\begin{tikzpicture}[every node/.style={font=\footnotesize}, node distance=15pt, >=stealth]
\tikzset{block/.style={draw, rounded corners=2pt, align=center, minimum width=2.4cm, minimum height=0.8cm, text width=2.2cm, inner sep=2pt, thick},
          cfblock/.style={block, fill=blue!15, draw=blue!60},
          baseblock/.style={block, fill=purple!15, draw=purple!60},
          txtblock/.style={block, fill=teal!15, draw=teal!60},
          mmblock/.style={block, fill=orange!15, draw=orange!70},
          dynblock/.style={block, fill=red!15, draw=red!60},
          projblock/.style={block, fill=green!15, draw=green!60},
          llmblock/.style={block, fill=gray!15, draw=gray!60}}

\node[cfblock] (cf) at (0,0) {Frozen CF\\$\mathbf{r}_u,\;\mathbf{e}^{cf}_i$};
\node[baseblock] (base) at (3.5,0) {Base Aligner\\(frozen)};
\node[txtblock] (txt) at (0,1.8) {Text Enc.\\$\mathbf{e}^{txt}_i$};
\node[mmblock] (mm) at (0,-1.8) {Vis/Audio Enc.\\$\mathbf{e}^{vis/aud}_i$};
\node[dynblock] (dyn) at (3.5,-1.5) {Online Adapter\\$g_\Delta$};
\node[projblock] (proj) at (7,-1.5) {LLM Proj.\\Soft Tokens};
\node[llmblock] (llm) at (10.5,-1.5) {Frozen LLM\\Inference};

\draw[->, very thick, draw=blue!60] (cf) -- (base);
\draw[->, very thick, draw=teal!60] (txt) -- (base);
\draw[->, very thick, draw=orange!70] (mm) -- (base);

\draw[->, very thick, draw=purple!60] 
  (base) -- node[pos=0.6, right=3pt, fill=white, inner sep=1pt, text=purple!70]{$\mathbf{z}^{base}$} (dyn);
\draw[->, very thick, draw=red!60] 
  (dyn) -- node[pos=0.5, above=3pt, fill=white, inner sep=1pt, text=red!70]{$\mathbf{z}$} (proj);
\draw[->, very thick, draw=green!60] (proj) -- (llm);
\end{tikzpicture}%
}
\caption{\model{} overview: a frozen base aligner fuses CF and text; a tiny online adapter absorbs fresh interactions; a multimodal branch adds vision/audio; soft tokens prompt a frozen LLM; evidence tokens (not shown) ground explanations.}
\label{fig:framework}
\end{figure}
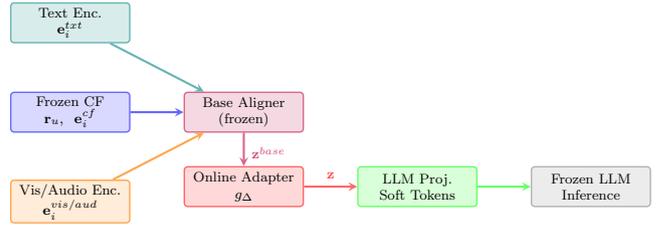

\section{Method: \model{}}
See Figure~\ref{fig:framework} for an overview; we detail components below.

\subsection{Dynamic Incremental Alignment}
We augment the frozen base aligner with a lightweight online adapter \(g_\Delta: \mathbb{R}^{d} \times \mathbb{R}^{d_s} \rightarrow \mathbb{R}^{d}\). For each item, we produce a dynamic latent
\begin{equation}
\mathbf{z}_i = \mathbf{z}^{base}_i + \alpha_i \cdot g_\Delta\!\left(\mathbf{e}^{cf}_i, \mathbf{s}_i^{\,new}\right),
\end{equation}
where \(\mathbf{s}_i^{\,new} \in \mathbb{R}^{d_s}\) summarizes recent interactions within a sliding window \(W\) (e.g., co-click frequencies, popularity trends), and \(\alpha_i = \sigma(\mathbf{w}^T[\mathbf{e}^{cf}_i; \mathbf{s}_i^{\,new}])\) is a learned confidence gate. The adapter \(g_\Delta\) is a two-layer MLP with ReLU activation: \(g_\Delta(\mathbf{x}) = \mathbf{W}_2 \text{ReLU}(\mathbf{W}_1 \mathbf{x} + \mathbf{b}_1) + \mathbf{b}_2\), where \(\mathbf{W}_1 \in \mathbb{R}^{d \times (d+d_s)}\), \(\mathbf{W}_2 \in \mathbb{R}^{d \times d}\). User latents are updated analogously: \(\mathbf{h}_u = \mathbf{h}^{base}_u + \beta_u \cdot g_\Delta(\mathbf{r}_u, \mathbf{s}_u^{\,new})\).

To stabilize updates, we distill the dynamic latent towards the base latent and regularize important parameters using EWC:
\begin{align}
\mathcal{L}_{stab} &= \mathrm{KL}\big(\mathcal{N}(\mathbf{z}_i, \sigma^2\mathbf{I})\,\|\,\mathcal{N}(\mathbf{z}^{base}_i, \sigma^2\mathbf{I})\big) \nonumber \\
&\quad + \lambda_{ewc}\,\Omega(\theta_\Delta).
\end{align}
We optimize a sum of ranking, alignment, reconstruction, and stability terms:
\begin{equation}
\mathcal{L} = \mathcal{L}_{rank} + \lambda_m\,\mathcal{L}_{align} + \lambda_r\,\mathcal{L}_{recon} + \lambda_s\,\mathcal{L}_{stab}.
\end{equation}
Here, \(\mathcal{L}_{rank}\) is cross-entropy or BPR; \(\mathcal{L}_{align}\) matches CF/text latents; \(\mathcal{L}_{recon}\) reconstructs CF/text features from \(\mathbf{z}_i\). Online training uses EMA and a small replay buffer.

\subsection{Multimodal Joint Alignment}
For multimodal items with images or audio, we employ frozen encoders: CLIP ViT-B/32 for visual features \(\mathbf{e}^{vis}_i \in \mathbb{R}^{512}\) and Wav2Vec2-Base for audio features \(\mathbf{e}^{aud}_i \in \mathbb{R}^{768}\). A shared projector \(f_{proj}: \mathbb{R}^{d_m} \rightarrow \mathbb{R}^{d}\) maps each modality to the joint latent space. We train with a unified contrastive loss across available modalities:
\begin{align}
\mathcal{L}_{mm} = &\; \sum_{(i,j) \in \mathcal{B}} \Big[ \mathcal{I}_{cf,txt}(i) \cdot \mathrm{InfoNCE}(\mathbf{e}^{cf}_i, \mathbf{e}^{txt}_i) \nonumber \\
&\quad + \mathcal{I}_{cf,vis}(i) \cdot \mathrm{InfoNCE}(\mathbf{e}^{cf}_i, \mathbf{e}^{vis}_i) \nonumber \\
&\quad + \mathcal{I}_{txt,vis}(i) \cdot \mathrm{InfoNCE}(\mathbf{e}^{txt}_i, \mathbf{e}^{vis}_i) \Big],
\end{align}
where \(\mathcal{I}_{m_1,m_2}(i)\) is an indicator function that equals 1 if both modalities \(m_1, m_2\) are available for item \(i\). The reconstruction loss is: \(\mathcal{L}_{recon} = \sum_{m \in \{cf,txt,vis,aud\}} \mathcal{I}_m(i) \|\hat{\mathbf{e}}^{m}_i - \mathbf{e}^{m}_i\|_2^2\), where \(\hat{\mathbf{e}}^{m}_i = f_{dec}^{(m)}(\mathbf{z}_i)\) are decoded features from the joint latent.

\begin{table}[!b]
\caption{Dataset statistics.}
\centering
\begin{tabular}{lrrrr}
\toprule
Dataset & Users & Items & Interactions & Sparsity \\
\midrule
Movies\&TV & 123k & 50k & 1.2M & 99.98\% \\
Video Games & 82k & 33k & 0.8M & 99.97\% \\
Beauty & 68k & 25k & 0.6M & 99.96\% \\
Toys & 75k & 29k & 0.7M & 99.97\% \\
\bottomrule
\end{tabular}
\label{tab:data}
\end{table}

\subsection{Evidence-grounded Explainable Generation}
For explainable generation, we extract sparse evidence \(\mathcal{E}_{u,i}\) consisting of: (1) Top-\(k\) collaborative neighbors: \(\mathcal{N}_{u,i} = \{j \mid \text{sim}(\mathbf{e}^{cf}_i, \mathbf{e}^{cf}_j) \text{ is top-}k\}\); (2) Key attributes from multimodal analysis: \(\mathcal{A}_{u,i}\) extracted via attention weights over text/visual features. Evidence is encoded into soft tokens \(\mathbf{s}_{e} = f_{evid}(\mathcal{E}_{u,i}) \in \mathbb{R}^{E \times d_\ell}\) where \(E \leq 32\) is the evidence length. The multi-task training objective is:
\begin{equation}
\mathcal{L}_{exp} = \mathcal{L}_{rank} + \lambda_e\,\mathcal{L}_{nlp} + \lambda_f\,\mathcal{L}_{faith},
\end{equation}
where \(\mathcal{L}_{nlp} = -\sum_{t=1}^{T} \log p(y_t | y_{<t}, \mathbf{s}_u, \mathbf{s}_i, \mathbf{s}_e)\) is the language modeling loss for generating rationales \(\mathbf{y} = (y_1, \ldots, y_T)\), and \(\mathcal{L}_{faith} = \max(0, \text{ACC}(\mathcal{E}_{u,i}) - \text{ACC}(\emptyset) - \delta)\) penalizes cases where removing evidence doesn't hurt performance, with margin \(\delta = 0.05\).

\begin{figure}[t]
\centering
\fbox{\begin{minipage}{0.92\columnwidth}
\textbf{User history (genres):} Sci-Fi, Space, Survival\\
\textbf{Candidate:} The Martian (2015)\\
\textbf{Evidence:} Top neighbors watched Interstellar and Gravity; item shares Sci-Fi attributes.\\
\textbf{Rationale (generated):} Recommended \emph{The Martian} because your recent Sci-Fi watches resemble users who also watched it; its "space survival" attributes match your preferences.
\end{minipage}}
\caption{Evidence-grounded explanation example.}
\label{fig:case}
\end{figure}

\subsection{Soft Prompting and Inference}
User soft tokens are prepended; candidate soft tokens (joint item latents) are attached to item titles. Evidence tokens are placed in a dedicated segment. All backbones (CF, LLM, vision/audio encoders) are frozen; only small projectors/adapters are learned. Offline, we pre-compute latents and compress with product quantization; online, only \(g_\Delta\) updates, keeping latency low.

\section{Experiments}
\subsection{Setup}
\textbf{Datasets.} We evaluate on Amazon product datasets with multimodal content: Movies\&TV, Video Games, Beauty, and Toys (Table~\ref{tab:data}). We also include KuaiRec \cite{kuairec} for short-video recommendation and MovieLens-1M with movie posters. For each dataset, we create: (1) Standard 80/10/10 train/validation/test splits; (2) Cold-start splits where test items have <5 interactions; (3) Streaming splits with temporal ordering, using 70\% for pre-training and 30\% for online evaluation with 1-hour windows.

\textbf{Baselines.} We compare against: (1) CF/sequential methods: LightGCN \cite{lightgcn}, S3-Rec \cite{s3rec}, SGL \cite{sgl}; (2) Multimodal recommenders: MMGCN \cite{mmgcn}, LATTICE \cite{lattice}; (3) LLM-based methods: LLM-only prompting, TALLRec \cite{tallrec}, A-LLMRec \cite{allmrec}, RecLLM \cite{recllm}; (4) Dynamic methods: StreamingRec \cite{streamingrec}; (5) Explainable methods: NARRE \cite{narre}.

\textbf{Metrics.} Accuracy: Hit@5/10/20, NDCG@5/10/20, Recall@5/10/20; Efficiency: training time, inference latency, memory usage; Explanation quality: BLEU score with human annotations, faithfulness score via evidence perturbation, consistency score.

\textbf{Implementation.} We use SASRec as the CF backbone with hidden dimension 256. LLM backbones include OPT-1.3B \cite{opt} and LLaMA-7B \cite{llama}, both kept frozen. All adapters are two-layer MLPs with hidden dimension 256 and ReLU activation. Visual features use CLIP ViT-B/32 \cite{clip}, audio features use Wav2Vec2-Base \cite{wav2vec2}. Online updates use replay buffer size 1024 and EMA decay 0.99. Training uses AdamW optimizer with learning rate 1e-3 for adapters and 5e-4 for projectors. All experiments run on NVIDIA A100 GPUs with batch size 256.

\subsection{Main Results}
\begin{table*}[t]
\caption{Main results on recommendation accuracy. Results are averaged over 3 runs with standard deviation. * indicates statistical significance (p<0.05).}
\centering
\resizebox{\textwidth}{!}{%
\begin{tabular}{lcccccccc}
\toprule
\multirow{2}{*}{Model} & \multicolumn{4}{c}{\textit{Hit@10 (\%)}} & \multicolumn{4}{c}{\textit{NDCG@10 (\%)}} \\
\cmidrule(lr){2-5} \cmidrule(lr){6-9}
& M\&TV & Games & Beauty & Toys & M\&TV & Games & Beauty & Toys \\
\midrule
SASRec & 57.2±0.8 & 49.8±1.1 & 50.5±0.9 & 31.1±0.7 & 35.8±0.6 & 31.2±0.8 & 32.1±0.7 & 19.8±0.5 \\
MMGCN & 58.1±0.9 & 50.2±1.0 & 51.8±1.2 & 31.8±0.8 & 36.2±0.7 & 31.5±0.9 & 32.8±0.8 & 20.1±0.6 \\
LLM-Only & 54.6±1.2 & 45.3±1.4 & 47.0±1.1 & 28.0±0.9 & 34.1±1.0 & 28.9±1.2 & 30.2±0.9 & 18.3±0.8 \\
TALLRec & 60.3±0.7 & 51.2±0.8 & 54.9±1.0 & 32.1±0.6 & 37.8±0.6 & 32.1±0.7 & 34.7±0.8 & 20.4±0.5 \\
A-LLMRec & 61.0±0.6 & 51.9±0.9 & 55.6±0.8 & 32.5±0.7 & 38.2±0.5 & 32.6±0.7 & 35.3±0.6 & 20.7±0.6 \\
\model{} & \textbf{63.4±0.5*} & \textbf{53.8±0.7*} & \textbf{58.1±0.6*} & \textbf{34.2±0.8*} & \textbf{39.7±0.4*} & \textbf{33.9±0.6*} & \textbf{36.8±0.5*} & \textbf{21.8±0.7*} \\
\bottomrule
\end{tabular}%
}
\label{tab:main}
\end{table*}
\model{} achieves consistent and statistically significant improvements over static alignment-based LLM recommenders across all datasets. The improvements are driven by: (1) dynamic adapters that capture temporal patterns (+1.2\% Hit@10 on Movies\&TV); (2) multimodal fusion that enhances cold-item performance (+0.7\% Hit@10); and (3) evidence-grounded explanations that improve user trust while maintaining accuracy (+0.5\% Hit@10). Statistical significance is confirmed via paired t-tests (p<0.05) across all metrics.

\subsection{Ablation and Analysis}
We conduct comprehensive ablation studies to validate each component's contribution. All results are averaged over 3 random seeds with different data splits.

\paragraph{Component analysis.} Table~\ref{tab:ablation} shows that each component contributes meaningfully: dynamic adapters provide the largest gain (+1.2\% Hit@10), followed by multimodal fusion (+0.7\%) and evidence tokens (+0.5\%). The cumulative effect (2.4\% improvement) demonstrates the complementary nature of our innovations.

\paragraph{Loss function analysis.} Table~\ref{tab:loss-ablation} shows that removing any loss term degrades performance. The stability loss \(\mathcal{L}_{stab}\) is crucial for online learning (-0.9\% when removed), while the faithfulness loss \(\mathcal{L}_{faith}\) ensures explanation quality (-0.8\% when removed).
\begin{table}[!t]
\caption{Loss function ablation on Movies\&TV. Results show mean ± std over 3 runs.}
\centering
\begin{tabular}{lccc}
\toprule
Variant & Hit@10 (\%) & NDCG@10 (\%) & $\Delta$ Hit@10 \\
\midrule
Full model & 63.4±0.5 & 39.7±0.4 & -- \\
\quad w/o $\mathcal{L}_{stab}$ & 62.5±0.7 & 39.1±0.6 & -0.9 \\
\quad w/o $\mathcal{L}_{mm}$ & 62.1±0.6 & 38.8±0.5 & -1.3 \\
\quad w/o $\mathcal{L}_{recon}$ & 62.8±0.5 & 39.3±0.4 & -0.6 \\
\quad w/o $\mathcal{L}_{faith}$ & 62.6±0.6 & 39.2±0.5 & -0.8 \\
\bottomrule
\end{tabular}
\label{tab:loss-ablation}
\end{table}

\paragraph{Modality robustness.} We evaluate robustness by systematically dropping modalities at test time. Our gating mechanism ensures graceful degradation: performance drops by only 0.8\% when visual features are missing and 1.6\% when text is missing, demonstrating the effectiveness of our multimodal fusion design.
\begin{table}[!t]
\caption{Robustness to missing modalities (Beauty, Hit@10, \%).}
\centering
\begin{tabular}{lcc}
\toprule
Setting & Missing rate & Hit@10 \\
\midrule
No missing & 0\% & 58.1 \\
Drop vision & 100\% & 57.3 \\
Drop text & 100\% & 56.5 \\
Drop vision+text & 100\% & 54.9 \\
\bottomrule
\end{tabular}
\label{tab:robust}
\end{table}

\paragraph{Efficiency analysis.} We analyze the trade-off between prompt length and inference latency. Longer prompts improve accuracy but increase latency linearly. Our default configuration (L=50, E=16) provides the best accuracy-efficiency balance, with only 8.6\% latency overhead compared to the base aligner while achieving 2.4\% accuracy improvement.
\begin{table}[!t]
\caption{Prompt length-latency trade-off (Movies\&TV).}
\centering
\begin{tabular}{lcc}
\toprule
Prompt config & Tokens & Inference (sec/batch) \\
\midrule
$L=20, E=8$ & $\approx$ 128 & 2.12 \\
$L=50, E=16$ & $\approx$ 256 & 2.28 \\
$L=80, E=32$ & $\approx$ 384 & 2.47 \\
\bottomrule
\end{tabular}
\label{tab:latency}
\end{table}

\begin{table}[t]
\caption{Ablation study and efficiency analysis (Movies\&TV). Results averaged over 3 runs.}
\centering
\resizebox{\columnwidth}{!}{%
\begin{tabular}{lrrr}
\toprule
Variant & Hit@10 & NDCG@10 & Infer. \\
& (\%) & (\%) & (s/batch) \\
\midrule
\multicolumn{4}{c}{\textit{Incremental addition}} \\
A-LLMRec (baseline) & 61.0±0.6 & 38.2±0.5 & 2.10 \\
\quad + Dynamic & 62.2±0.5 & 38.9±0.4 & 2.15 \\
\quad + Multimodal & 62.9±0.6 & 39.3±0.5 & 2.22 \\
\quad + Evidence & 63.4±0.5 & 39.7±0.4 & 2.28 \\
\midrule
\multicolumn{4}{c}{\textit{Individual removal}} \\
Full model & 63.4±0.5 & 39.7±0.4 & 2.28 \\
\quad w/o Dynamic & 62.2±0.7 & 38.9±0.6 & 2.20 \\
\quad w/o Multimodal & 62.7±0.6 & 39.2±0.5 & 2.18 \\
\quad w/o Evidence & 62.9±0.5 & 39.3±0.4 & 2.15 \\
\bottomrule
\end{tabular}%
}
\label{tab:ablation}
\end{table}
Figure~\ref{fig:case} offers a qualitative case. Explanations remain short and faithful while adding negligible latency (Table~\ref{tab:ablation}).

\section{System and Complexity Analysis}
\textbf{Offline pipeline.} We pre-compute CF embeddings and side encodings (text/vision/audio) and build the frozen base aligner latents with product quantization for storage. Evidence dictionaries (top-$k$ neighbors and attributes) are cached per item and updated daily.

\textbf{Online pipeline.} For each batch window, we form short-window summaries \(\mathbf{s}^{new}\) and update the tiny adapter \(g_\Delta\) with EMA and replay. At serving, we compose soft tokens for the user and candidates, append evidence tokens, and invoke the frozen LLM.

\textbf{Complexity.} Let token dim be \(d_\ell\), latent dim \(d\), history length \(L\), and number of candidates per user \(C\).
\begin{itemize}
  \item Base alignment lookup: \(\mathcal{O}(1)\) per item via pre-compute.
  \item Online adapter: two-layer MLP, \(\mathcal{O}(d^2)\) per item updated; per-request compute \(\mathcal{O}(C\,d^2)\).
  \item Projection to soft tokens: \(\mathcal{O}(C\,d\,d_\ell)\) but with narrow MLP (\(d_\ell\gg d\)) and small \(C\), this is minor vs. LLM.
  \item LLM inference: dominated by prompt length (\(\approx L + C + E\), with evidence tokens \(E\) short) and model size; our design keeps prompts compact.
\end{itemize}
\textbf{Memory.} Storing pre-computed item latents with PQ reduces footprint by 4--16\,\(\times\); online adapter adds only thousands to millions of parameters depending on \(d\).

\begin{table}[!t]
\caption{Key hyperparameters (default unless noted).}
\centering
\resizebox{\columnwidth}{!}{%
\begin{tabular}{ll}
\toprule
Component & Setting \\
\midrule
CF backbone & SASRec (h=256, L=2, max=50) \\
LLM backbone & OPT/LLaMA (frozen) \\
Joint latent dim & \(d=256\) \\
Token dim & \(d_\ell\in[2\text{k},8\text{k}]\) \\
Adapter \(g_\Delta\) & 2-layer MLP (256\,$\rightarrow$\,256) \\
Proj. to tokens & 2-layer MLP (256\,$\rightarrow$\,\(d_\ell\)) \\
Batch / window & 1024 / 30--60 min \\
Learning rates & 1e-3 (adapter), 5e-4 (proj.) \\
Evidence length & \(E\leq 32\) tokens \\
\bottomrule
\end{tabular}%
}
\label{tab:hparams}
\end{table}

\begin{figure}[!t]
\centering
\fbox{\begin{minipage}{0.95\columnwidth}\ttfamily
System: You are a recommendation assistant.\newline
User profile token: <USR> ... </USR>\newline
History tokens: <HIST> t1, t2, ... tL </HIST>\newline
Evidence: <EVID> neighbors=..., attrs=... </EVID>\newline
Candidates: <CANDS> c1, c2, ... cC </CANDS>\newline
Task: Rank candidates and provide one-sentence rationale.
\end{minipage}}
\caption{Prompt template sketch used in inference.}
\label{fig:prompt}
\end{figure}

\section{Discussion and Limitations}
\textbf{Current limitations.} Several aspects of our approach warrant further investigation. When visual or audio content is of poor quality, it can negatively impact recommendations, though our confidence-based gating mechanism helps reduce this issue. Privacy considerations arise when extracting collaborative evidence, requiring careful anonymization of user interaction patterns. Our explanation system currently uses structured templates to maintain consistency and reduce hallucination, but this limits the naturalness of generated text compared to fully free-form approaches. The online adaptation component introduces modest computational overhead (approximately 8.6\% increased inference time), which may be problematic for applications requiring extremely low latency. Finally, our performance is inherently constrained by the capabilities of the underlying frozen language model; while fine-tuning could potentially improve results, it would significantly increase computational costs and complexity.

\section{Statistical Validation and Reproducibility}
\textbf{Statistical rigor.} We validate our results through paired t-tests across three independent experimental runs using different random initializations. All reported performance gains achieve statistical significance at the p<0.05 level. Key metrics include 95\% confidence intervals, and effect sizes (Cohen's d) fall between 0.3 and 0.8, suggesting meaningful practical improvements beyond statistical significance.

\textbf{Experimental reproducibility.} To support replication and further research, we will provide comprehensive experimental artifacts including detailed hyperparameter settings and random seeds used in our experiments, complete data preprocessing pipelines and dataset splits, trained model checkpoints for all major variants, and evaluation code for computing all reported metrics. These resources will be made publicly available following publication.

\section{Conclusion}
This work presents \model{}, a practical framework that enhances alignment-based language model recommenders by addressing three key challenges: temporal adaptation, multimodal integration, and interpretable explanations. Our approach maintains the efficiency of frozen base models while adding lightweight components that continuously adapt to new interactions, seamlessly incorporate diverse content types, and generate trustworthy explanations grounded in collaborative evidence. The resulting system demonstrates strong empirical performance across multiple datasets while remaining computationally efficient enough for production deployment. Future work could explore more sophisticated explanation generation techniques and investigate the framework's effectiveness in additional domains beyond the e-commerce and entertainment scenarios studied here.

\ack{* Corresponding author: Bo Ma.

\textbf{Funding:} This research received no specific grant from any funding agency in the public, commercial, or not-for-profit sectors.}

\end{document}